\documentclass[aps,twocolumn,superscriptaddress,preprintnumbers,showpacs] {revtex4}
\usepackage{epsfig,graphicx,times}
\usepackage{amsmath,amssymb,graphicx}
\usepackage{epsfig}
\usepackage{graphicx}
\usepackage{enumerate}

\newcommand{\beq}{\begin{equation}}
\newcommand{\eeq}{\end{equation}}

\newcommand{\bea}{\begin{eqnarray}}
\newcommand{\eea}{\end{eqnarray}}

\begin{document}


\title
{Exact dynamics of interacting qubits in a thermal environment:\\
Results beyond the weak coupling limit}
\author{Lian-Ao Wu}
\affiliation{Department of Theoretical Physics and History of
Science, The Basque Country University (EHU/UPV) and IKERBASQUE --
Basque Foundation for Science, 48011, Bilbao, Spain}
\author{Claire X. Yu}
\affiliation{Chemical Physics Theory Group, Department of Chemistry and Center for Quantum
Information and Quantum Control, University of Toronto, 80 St. George
street, Toronto, Ontario, M5S 3H6, Canada}
\author{Dvira Segal}
\affiliation{Chemical Physics Theory Group, Department of Chemistry and
Center for Quantum Information and Quantum Control, University of
Toronto, 80 St. George street, Toronto, Ontario, M5S 3H6, Canada}

\date{\today}

\begin{abstract}
We demonstrate an exact mapping of a class of models of two
interacting qubits in thermal reservoirs to two separate spin-bath
problems. Based on this mapping, exact numerical simulations of the
qubits dynamics can be performed, beyond the weak system-bath
coupling limit. Given the time evolution of the system, we study, in
a numerically exact way, the dynamics of entanglement between pair
of qubits immersed in boson thermal baths, showing a rich
phenomenology, including an intermediate oscillatory behavior, the
entanglement sudden birth, sudden death, and revival. We find that
stationary entanglement develops between the qubits due to their
coupling to a thermal environment, unlike the isolated qubits case
in which the entanglement oscillates. We also show that the
occurrence of entanglement sudden death in this model depends on the
portion of the zero and double excitation states in the subsystem
initial state. In the long-time limit, analytic expressions are
presented at weak system-bath coupling, for a range of relevant
qubit parameters.

\end{abstract}

\maketitle


\section{Introduction}

Understanding the dynamics of a dissipative quantum system is a
prominent challenge in physics, as a quantum system is never
perfectly isolated from a larger environment. A minimal, yet highly
rich model for exploring quantum dissipation effects, is the
spin-boson model, including an impurity two-level system (referred
to as a spin) coupled to a thermal reservoir. This model displays a
rich phase diagram in the equilibrium regime \cite{Legget,LehurR}.
The non-equilibrium version of this model, referring to the case
where the spin is coupled to two thermal reservoirs, has been
suggested as a toy model for exploring quantum transport
phenomenology through an anharmonic nanojunction \cite{Rectif,
SegalM}. In this case, the generic situation is one of a
non-equilibrium steady-state, regardless of the initial preparation.

\begin{figure}[htbp]
\hspace{-1mm} {\hbox{\epsfxsize=120mm \hspace{3mm}\epsffile{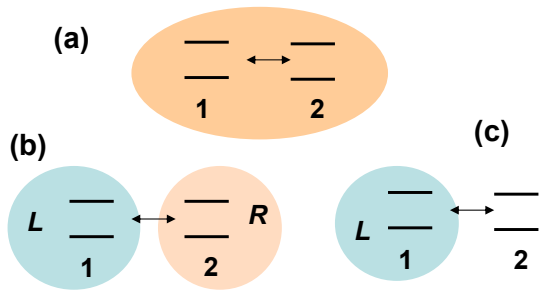}}}
\vspace{-115mm} \caption{Scheme of the model system including two
interacting qubits (a) immersed in a common bath, (b) coupled to
separate baths, $L$ and $R$, and (c) with qubit '2'  coupled to a
thermal bath only through its interaction with qubit '1'. This model
can represent the nonlinear coupling of qubit '2' to a structured
bath. Simulations were performed here assuming scenario (c).}
\label{FigS}
\end{figure}

Interacting two-level systems are the basic element in quantum
computation, thus it is paramount to extent the minimal spin-boson
scenario and describe more complex modular systems, e.g., two
interacting qubits immersed in a thermal environment \cite{Lehurq}.
For a schematic representation, see Fig. \ref{FigS}. The qubits may
share their thermal environment, or may separately couple to
independent baths, maintained at a nonzero temperature. The latter
situation corresponds to the case where the qubits are not
necessarily placed close to each other. In another relevant setup,
one qubit couples indirectly to a thermal reservoir, through its
interaction with the other qubit. This situation effectively
corresponds to a subsystem anharmonically coupled to a harmonic
bath, allowing to introduce nontrivial nonlinear effects
\cite{Grifoni}. Physical realizations include, for example,
ultracold atoms in optical lattices \cite{Cold}, trapped ions
\cite{Ion}, resonator-coupled superconducting qubit arrays
\cite{Josephson,Super}, and electron spins in quantum dots and doped
semiconductors \cite{QD}. In such systems one should consider (at
least) four energy scales: the internal qubit energetics,
controlling its isolated (Rabi oscillation) dynamics, qubit-bath
interaction strength and the environment temperature, leading to
decoherence and relaxation processes, and qubit-qubit coupling
energy, admitting state transfer between qubits and  a nontrivial
gate-functionality.

The dissipative multi-qubit system has recently served as a simple
model for resolving issues related to coherence dynamics in the time
evolution of biological molecules, e.g., the Fenna- Matthews-Olson
(FMO) complexes, resulting in the identification of the relevant
decoherence, relaxation and disentanglement timescales
\cite{Whaley,Thorwart, Pachon}. It should be noted that these works
have considered single-excitation states only, ignoring the
contribution of the zero and the doubly excited states in the
two-qubits dynamics.

In this paper, we analytically demonstrate that a class of
interacting two-qubit systems immersed in separate thermal
reservoirs or within a common bath, can be mapped onto two uncoupled
spin-bath problems, allowing for an exact numerical solution of the
qubits dynamics. For a bosonic environment and a particular
system-bath interaction form, we perform those simulations using an
exact numerical technique, the quasi adiabatic path-integral (QUAPI)
approach \cite{QUAPI}, providing the population and coherence
dynamics of the system. With this at hand, we can follow the exact
dynamics of entanglement between the qubits, as quantified by
Wootters' concurrence \cite{Wooters}. For a range of system and bath
parameters, analytic results are presented, describing the system
behavior in the long time limit. The model investigated here is more
general than what has been typically considered before, going beyond
the simple exchange interaction model \cite{Thorwart,Pachon}.

Entanglement is associated with nonclassical correlations between
two or more quantum systems \cite{EntagR}. Since it is a
basic resource in quantum computation and information technology, it
is important to understand the extent to which environmental-induced
decoherence processes degrade and destroy it \cite{TingYu}, or
alternatively, generate \cite{Benatti} and maintain it
\cite{ShiLiang}. It has been recently shown that two qubits in
separate reservoirs may disentangle {\it at finite times}, as
opposed to the behavior of coherences. This process is referred to
as ``entanglement sudden death" \cite{TingYu,DeathExp}. More recent
theoretical and experimental studies have looked at related effects,
e.g., the collapse and subsequent revival of the entanglement
\cite{Guo}, or its delayed-sudden birth, induced by a dissipative
bath \cite{Tanas}. Steady-state entanglement generation by
dissipation has been recently observed in atomic ensembles
\cite{Ciracss}. These studies have assumed non-interacting qubits,
and the system dynamics has been typically followed within quantum
master equation approaches (e.g., the Redfield equation or Lindblad
formalism \cite{Petrucci}), by invoking the weak system-bath
coupling approximation. The markovian limit has been further assumed
in many cases, see e.g., \cite{Benatti,Asma}.

Our work here departs from these studies in two substantial aspects.
First, we consider a more complex model for the subsystem,
introducing qubit-qubit interaction, with the motivation to examine
a setup relevant for quantum computing technology. Using an exact
mapping, we show how the dynamics can be followed within a simpler
construction: While we take into account the zero excitation and
double excitation states, under certain initial conditions their
dynamics can be separated from the evolution of the
single-excitation states. However, their contribution to the pair
entanglement is paramount. Second, we refrain from making
approximations: We study the qubits dynamics using a numerically
exact method, assuming a class of initial states. The results are
valid beyond the weak system-bath coupling scenario, accommodating
non-markovian effects.

Our calculations display rich dynamics. Particularly, we observe the
development of a {\it stationary} concurrence due to the coupling of
the qubits to thermal baths. This result stands in a direct contrast
to the oscillatory behavior observed in the fully coherent regime.
It demonstrates that while entanglement inherently relies on the
existence of quantum correlations in the system, it nevertheless
requires non-vanishing decoherence and relaxation effects in order
to be stabilized and  become useful for quantum technologies. Other
phenomena detected and explained here are entanglement delayed
sudden birth, sudden death, and revival.

The paper is organized as follows. In Sec. II we describe the model
of interest, and explain its mapping onto two separate spin-bath
problems. In Sec. III we explain how we follow the system dynamics,
and include relevant expressions for calculating the qubits
concurrence. Numerical results within QUAPI are included in Sec. IV.
The long-time limit is  discussed in Sec. V. Sec. VI concludes.


\section{Model}

The general model to be considered here includes
two interacting qubits, $i=1,2$, 
immersed within separate reservoirs, $L$ and $R$, respectively. The
formalism can be reduced to describe a single-bath scenario. The
total Hamiltonian includes three terms,
\bea H=H_{S}+H_{B}+V_{SB}. \label{eq:HT} \eea
$H_{S}$ and $H_{B}=H_L+H_R$ stand for the system and reservoirs
Hamiltonians, respectively. The former includes the isolated qubits
with the internal energy bias $\epsilon_{i}$ and a qubit-qubit
interaction term $V_{ss}$,
\bea
H_{S}&=&\epsilon _{1}\sigma _{1}^{z}+\epsilon _{2}\sigma _{2}^{z}+V_{ss},
\nonumber\\
V_{ss}&=&\frac{J}{2}[(1+\gamma )\sigma _{1}^{x}\sigma _{2}^{x}+(1-\gamma
)\sigma _{1}^{y}\sigma _{2}^{y}+2\delta \sigma _{1}^{z}\sigma _{2}^{z}].
\eea
$\sigma_i^{p}$ ($p=x,y,z$) are the Pauli matrices for the $i$th
spin, $J$ is an energy parameter characterizing the exchange
interaction, $\gamma$ and $\delta$ set the interaction anisotropy.
Our mapping holds for a dephasing-type system-bath interaction model,
%
\bea
V_{SB}=\sigma _{1}^{z}B_{L}+\sigma _{2}^{z}B_{R}.
\label{eq:deph}
\eea
Here, $B_{\nu}$ is a $\nu=L,R$ bath operator, with $B_L$ coupled to
spin '1' and $B_R$ coupled to spin '2'. In our simulations below we
adopt bosonic reservoirs:
each thermal baths includes a collection of independent harmonic oscillators,
\bea H_{\nu}=\sum_{k\in \nu}\omega_ka_k^{\dagger}a_k.\eea
The operators $a_k^{\dagger}$ and $a_k$ are bosonic creation and
annihilation operators, respectively, $\omega_k$ is the mode
frequency. We also assume that the interaction operators constitute the
reservoirs displacements from equilibrium,
\bea
B_{\nu}=\sum_{k\in
\nu} \lambda_k \left(a_k^{\dagger}+a_k\right).
\eea
Here, $\lambda_{k}$ are system-bath coupling constants. The mapping
described next, from the Hamiltonian (\ref{eq:HT}) into two
spin-bath problems, neither rely on a particular bath statistics nor
on the details of the operators $B_{\nu}$. For example, it is valid
for a model of two-qubits in fermionic or spin environments.
The Hamiltonian introduced so far takes into account two independent
reservoirs. We could explore a similar setup with one qubit coupled
to a thermal reservoir indirectly, through its interaction with the
second qubit,  mimicking nonlinear effects \cite{Grifoni}, see Fig. \ref{FigS}.
Another relevant setup includes two qubits immersed in a common
thermal reservoir. 

The Hilbert space of the qubits is spanned by four vectors,
$\left|00\right\rangle, \left|01\right\rangle,
\left|10\right\rangle, \left| 11\right\rangle$, forming the
excitation basis of the Hamiltonian, where the left (right) digit
indicates the state of qubit 1 (2). We now show that the Hamiltonian
can be mapped onto two spin-bath type models. We begin by defining
four composite system operators,
\bea P_{z}&=&\frac{1}{2}(\sigma_{1}^{z}-\sigma _{2}^{z}), \,\,\,
Q_{z}=\frac{1}{2}(\sigma _{1}^{z}+\sigma _{2}^{z}), \,\,\,
\nonumber \\
P_{x}&=&\frac{1}{2}\left( \sigma_{1}^{x}\sigma _{2}^{x}+\sigma
_{1}^{y}\sigma _{2}^{y} \right), \,\,\, Q_{x}=\frac{1}{2}
\left(\sigma _{1}^{x}\sigma _{2}^{x}-\sigma _{1}^{y}\sigma _{2}^{y}
\right). \eea
In the excitation basis, these operators take the explicit form
\bea P_x= \left|01\right\rangle\langle 01| +
\left|10\right\rangle\langle10|, \,\, P_z=
\left|10\right\rangle\langle 01| -  \left|01\right\rangle\langle10|
\nonumber\\
Q_x= \left|00\right\rangle\langle 11| +
\left|11\right\rangle\langle00|, \,\, Q_z=
\left|00\right\rangle\langle 00| -  \left|11\right\rangle\langle11|.
\eea
Additionally, we construct two identity-type operators,
\bea I_P&=&\left|10\right\rangle\langle 01| +
\left|01\right\rangle\langle10|
\nonumber\\
I_Q&=& \left|00\right\rangle\langle 00| +
\left|11\right\rangle\langle11|.
\eea
With these at hand, the Hamiltonian (\ref{eq:HT}) can be written
as
\bea
H=H_Q+H_P,
\eea
where
\bea
H_Q&=&
\epsilon Q_{z} + J\gamma Q_{x} + Q_{z}B+J\delta I_Q +H_B I_Q
\nonumber\\
H_P &=&\bar{\epsilon}P_{z}+JP_{x} +P_{z}\overline{B}- J\delta I_P
+H_{B}I_P. 
\label{eq:HTR}
\eea
Here, $\epsilon =\epsilon_{1}+\epsilon _{2}$,
$\bar{\epsilon}=\epsilon_{1}-\epsilon _{2}$, $B=B_{L}+B_{R}$ and
$\overline{B}=B_{L}-B_{R}$. One can easily show that the following
commutators vanish \cite{para} ($m,n=x,z$)
\bea \lbrack Q_{m},P_{n}]=[P_{m},\sigma _{1}^{z}\sigma
_{2}^{z}]=[Q_{m},\sigma _{1}^{z}\sigma _{2}^{z}]=0. \eea
The Hilbert space of two qubits can thus be factored into two
direct-sum subspaces: The first, $P$, is spanned by
$\left|01\right\rangle$ and $\left|10\right\rangle$. The second,
$Q$, is spanned by $\left|00\right\rangle$ and$\left|
11\right\rangle$. One can further prove that $P_{x},[P_{x},P_{z}]$
and $P_{z}$ generate an $SU^{P}(2)$ group, and $Q_{x},[Q_{x},Q_{z}]$
and $Q_{z}$ generate another $SU^{Q}(2)$ group. The two groups have
a direct-sum structure, $SU^{P}(2)\oplus SU^{Q}(2)$. We now note
that $P_{x}$ and $P_{z}$ in subspace $P$ play the role of the Pauli
matrices $\sigma^{x} $ and $\sigma^{z}$ (in the space spanned by
$\left| 0\right\rangle $ and $\left| 1\right\rangle$). The same
principle holds for $Q_{x}$ and $Q_{z}$ in the $Q$ subspace.
Overall, in mapping Eq. (\ref{eq:HT}) into Eq. (\ref{eq:HTR}) we
replaced a model of two interacting spins coupled each to its own
thermal reservoir, by a model of two separate spin-boson-type
systems, where each spin in the new model is coupled to {\it both}
reservoirs. The latter model is significantly simpler than the
former, and we can explore its dynamics using exact simulation tools
for a class of certain initial conditions.

The mapping described here holds for general reservoirs and a
bilinear system-bath interaction form, with an
arbitrary bath operator coupled to the subsystem. The
results also hold when we apply a dressing transformation
\cite{dress} $W$, $H^{\prime }=W^{\dagger }HW$ on the two qubits
Hamiltonian, or the reservoirs. For example, we may introduce a
spin-orbital coupling into $V_{ss}$ via $W=\exp (i\frac{\theta}{2} \sigma
_{1}^{z})$, while keeping other terms in the total Hamiltonian
unchanged
\bea V_{ss}^{\prime }=\cos \theta V_{ss}+\sin \theta
\frac{J}{2}[(1+\gamma )\sigma _{1}^{y}\sigma_{2}^{x}-(1-\gamma
)\sigma _{1}^{x}\sigma _{2}^{y}]. \eea
Another relevant case that can be simulated exactly relies on the
absence of external fields, $\epsilon _{1}=\epsilon _{2}=0$, taking
$W=\exp (i\frac{\pi }{4}(\sigma _{1}^{y}+\sigma _{2}^{y}))$. This
results in
\bea H_{S}^{\prime }=\frac{J}{2}[(1+\gamma )\sigma _{1}^{z}\sigma
_{2}^{z}+(1-\gamma )\sigma _{1}^{y}\sigma _{2}^{y}+2\delta \sigma
_{1}^{x}\sigma _{2}^{x}] \eea
with $V_{SB}'=\lambda _{1}\sigma_{1}^{x}B_{L}+\lambda_{2}\sigma
_{2}^{x}B_{R}$. After this transformation, the model has turned into
the anisotropic XYZ-type model with flip-flop ($\sigma_x$) coupling
between the system and reservoirs.
In this form, the model describes energy exchange
between the qubits and the baths,
unlike the original Hamiltonian [Eq. (\ref{eq:deph})]
which delineates dephasing effects.
When $1-\gamma = 2\delta$, it reduces
to the standard XY model, $H_{S}^{\prime }=\frac{J}{2}[(1-\gamma
)(\sigma _{1}^{y}\sigma _{2}^{y}+ \sigma_{1}^{x}\sigma
_{2}^{x})+(1+\gamma )\sigma _{1}^{z}\sigma_{2}^{z}] $.

\section{Dynamics and quantum entanglement}

We explain here how we time-evolve the reduced density matrix, to
obtain the qubits dynamics. As an initial condition we consider a
system-bath product state, where the reservoirs are maintained in a
canonical-thermal state,
$\rho_{\nu}=e^{-\beta_{\nu}H_{\nu}}/Z_{\nu}$, $Z_{\nu}={\rm
Tr_B}[e^{-\beta_{\nu}H_{\nu}}]$ is the partition function.
In order to separate the dynamics into the $Q$ and $P$ branches, we must
adopt a direct sum $Q$-$P$ initial state for the qubits. Overall,
the total density matrix at time $t=0$ is written as
 \bea \rho(0)= \left(
\begin{array}{cc}
\rho_P(0) & 0   \\
0 & \rho_Q(0)  \\
\end{array} \right) \otimes \rho_B.
\label{eq:t0} \eea
For a particular example, see Eq. (\ref{eq:init}). Under this
construction, the reduced density matrix follows
\bea &&\rho_S(t)={\rm Tr_B}[ U(t)\rho(0)U^{\dagger}(t)]
\nonumber\\
&&= {\rm Tr_B}\left( \begin{array}{cc}
U_P(t)\rho_P(0) \rho_B U^{\dagger}_P(t) & 0   \\
0 &    U_Q(t)\rho_Q(0) \rho_B U^{\dagger}_Q(t)  \\
\end{array} \right)
\label{eq:rhoS}
\nonumber \\
\eea
The trace is performed over the $L$ and $R$ degrees of freedom.
The time evolution operators, $U(t)=U_Q(t)\oplus U_P(t)$, are defined as
\bea U_{Q}(t) =e^{-itH_Q}, \,\,\,
U_{P}(t) =e^{-itH_P},
\label{eq:timeE}
\eea
with $H_P$ and $H_Q$ given in Eq. (\ref{eq:HTR}). Eq.
(\ref{eq:rhoS}) establishes an important result: The dynamics of the
two-qubit system proceeds in two independent branches, each
equivalent to a spin-bath model. In the case of bosonic baths, the
dynamics in each branch is followed next using the QUAPI technique
\cite{QUAPI}, a numerically exact simulation tool that can be easily
extended to include more than one thermal reservoir. The output of
this calculation is the reduced density matrix of the qubits. We use
this information and investigate the time evolution of entanglement
between the qubits. As a side comment, we note that if the qubits
were to couple to a common bath, $\bar B=0$, only the $Q$ subspace
would have become susceptible to decoherring effects, while the $P$
subspace would be an invariant subspace, or a ``decoherence free"
subspace.


Based on Eqs. (\ref{eq:t0})-(\ref{eq:timeE}), we conclude that the
reduced density matrix $\rho_S$  is an X-type matrix at all times,
once we organize it in the standard order of basis vectors
 $\left| 00\right\rangle ,\left| 01\right\rangle ,\left|
10\right\rangle \ $and $\left| 11\right\rangle$,
\bea &&\rho_S=
\left( \begin{array}{cccc}
(\rho_{S})_{00,00}& 0 & 0 & (\rho_{S})_{00,11}    \\
0& (\rho_{S})_{01,01}  & (\rho_{S})_{01,10}  & 0    \\
0& (\rho_{S})_{10,01}  & (\rho_{S})_{10,10}  & 0    \\
(\rho_{S})_{11,00}& 0 & 0 & (\rho_{S})_{11,11}    \\
\end{array} \right)
\label{eq:M}
\nonumber \\
\eea
This is an important result: The dynamics of a class of dissipative interacting qubits
[Eqs. (\ref{eq:HT})-(\ref{eq:deph})] can be reached via the solution
of two spin-bath
problems, and the reduced density matrix satisfies an X-form
at all times.
Different quantities are of interest, e.g., the timescale for maintaining
coherences in the system \cite{Pachon}.
Here, we focus on quantum correlations in the system \cite{EntagR},
computed next in a numerically exact way beyond weak coupling.
In particular, we quantify
the degree of entanglement between the qubits using Wootters'
concurrence \cite{Wooters}. For mixed states it is
calculated by considering the eigenvalues of the matrix
$r(t)=\rho_S(t)\sigma_1^{y}\otimes \sigma_2^{y}\rho_S ^{\ast
}(t)\sigma_1^{y}\otimes \sigma_2^{y}$,
given here by
\bea
\lambda_{1,2}
&=&\left[\sqrt{(\rho_S)_{01,01}(\rho_S)_{10,10}}\pm \left|
(\rho_S)_{01,10}\right| \right]^{2},
\nonumber\\
\lambda_{3,4}&=&\left[\sqrt{ (\rho_S)_{00,00}(\rho_S)_{11,11}}\pm \left| (\rho_S)_{00,11}\right|\right] ^{2}.
\eea
In terms of these eigenvalues, the concurrence is defined as
\bea
C(t)&=&\max (2\max (\sqrt{\lambda_{1}},\sqrt{\lambda_2}, \sqrt{\lambda_3}, \sqrt{\lambda_4}),
\nonumber\\
&-&\sqrt{\lambda_1}- \sqrt{\lambda_2}-\sqrt{\lambda_3}-\sqrt{\lambda_4},0).
\eea
It varies from $C=0$ for a disentangled state to $C=1$ for a maximally entangled state.
In the present case it reduces to
\bea
C(t)=\max(0, 2F_1,2F_2)
\label{eq:Ct}
\eea
with
\bea F_1&=& |(\rho_S)_{01,10}| -
[(\rho_S)_{00,00}(\rho_S)_{11,11}]^{1/2},
\nonumber\\
F_2&=& |(\rho_S)_{00,11}| - [(\rho_S)_{01,01}(\rho_S)_{10,10}]^{1/2}
\label{eq:Conc1}. \eea
The dynamics of concurrence for an $X$-state density matrix has been
examined in different works. For example, in Ref. \cite{Asma} it is
demonstrated that the effect of entanglement sudden death should
always take place in a noninteracting qubit system once coupled to a
finite temperature reservoir. As we mention in the introductory
section, we depart from this study and similar works in two aspects:
(i) We build the reduced density matrix using an exact numerical
treatment, and (ii) we consider a more general model, including
quit-qubit interaction effects, with the motivation to consider a
setup more relevant for quantum computation technologies.

\section{Numerical Results}

We simulate the spin-boson dynamics in the $P$ and $Q$ branches
(separately) using QUAPI \cite{QUAPI}, to obtain the population and
coherences in each branch. With this at hand, we generate the
$4\times 4$ reduced density matrix $\rho_S(t)$, Eq. (\ref{eq:M}).
The qubits degree of entanglement is calculated using Eq.
(\ref{eq:Ct}). Our general description assumes two thermal
reservoirs: $H_L$, coupled to spin 1 and $H_R$, coupled to spin 2.
These reservoirs are characterized by the spectral function
$J_{\nu}(\omega)=\pi \sum_{k\in \nu}
\lambda_{k}^2\delta(\epsilon_{k}-\omega)$. Specifically, we simulate
Ohmic baths, $J_{\nu}(\omega)=\frac{\pi K_{\nu}}{2}\omega
e^{-\omega/\omega_c}$; $\omega_c$ is the cutoff frequency. The
dimensionless prefactor $K_{\nu}$ is referred to as the Kondo
parameter, describing the strength of the system-bath interaction
energy. In practice, our simulations were performed without the $R$
reservoir, by taking $K_R=0$. The reason for this choice is that in
the spin-boson Hamiltonian (\ref{eq:HTR}) the inclusion of identical
reservoirs which are interacting in the same manner with the spins
(same functional form for $B_L$ and $B_R$, up to a sign), simply
amounts to an additive operation, reflecting a linear scaling of the
Kondo parameter when more than one reservoir is incorporated. In
what follows, we thus use the short notation $K\equiv K_{L}$,
$K_R=0$ and $T\equiv T_L$.

The energy parameters in the system are the qubit-qubit coupling,
taken as $J=1$, and the anisotropy parameters $\delta=0.1$,
$\gamma=0.5$. The qubits are assumed identical with
$\epsilon_1=\epsilon_2\sim 0.1-0.5$. For the reservoir we take as a
cutoff frequency $\omega_c=7.5$, and use temperatures at the range
$T=0.1-1$. The Kondo parameter extends from the weak coupling limit
($K=0.05$) to the strong-intermediate regime, $K=0.8$, where
convergence of QUAPI can be achieved. The following initial
condition is utilized    for the qubits subsystem
\bea
\rho_S(0)=a\rho_Q(0)\oplus(1-a)\rho_P(0),
\label{eq:Bell}
\eea
with $0\leq a\leq 1$ and $\rho_{Q,P}(0)$  as (maximally entangled) Bell states,
\bea \rho_Q(0)= \frac{1}{2}\begin{pmatrix} 1 & 1 \\ 1 & 1
\end{pmatrix}; \,\,\, \rho_P(0)=\frac{1}{2}\begin{pmatrix}  1 &1
\\ 1 & 1
\end{pmatrix}
\label{eq:init}
\eea
The concurrence (\ref{eq:Ct}) can be simplified if the following
conditions are simultaneously satisfied: (i)  $a\leq1/2$, and (ii)
$\epsilon_1=\epsilon_2$. 
The latter condition, combined with the initial state ascribing
identical weight to diagonal elements in the $P$ subspace, implies
that the populations in the $P$ (single excitation) subspace are
identical at all times,
$(\rho_S)_{01,01}=(\rho_S)_{10,10}=\frac{1-a}{2}$.
Since $\sqrt{(\rho_S)_{00,00}(\rho_S)_{11,11}}\leq \frac{a}{2}$ at
all times and the density matrix positivity condition demands that
$|\rho_{i,j}|^2\leq\rho_{i,i}\rho_{j,j}$, the off diagonal terms are
bounded by $|(\rho_S)_{11,00}|\leq \frac{a}{2}$. This implies that
$F_2$ cannot be larger than zero at any instant for $a\leq 1/2$. The
concurrence can then be simplified to
\bea C_{a\leq \frac{1}{2}}(t)=  \max (0,2F_1    ). \label{eq:Conc2}
\eea
This expression indicates that $C$ is nonzero if the magnitude of
the coherence in the $P$ subspace is large, in comparison to the
product of populations in the $Q$ subspace. Thus, to understand the
behavior of entanglement between the qubits  one needs to follow the
population and coherence dynamics in both subspaces. At the special
point $a=0$ the concurrence reduces to the simple form
\bea C_{a=0}(t)=\max \Big( |(\rho_S)_{01,10}| ,0 \Big),
\label{eq:Conc3} \eea
which only depends on coherences behavior, a continuous function. As
a result, concurrence sudden death is eliminated, indicating that
this effect is directly linked to the inclusion of zero and double
excitation components in the dynamics.


\begin{figure}[htbp]
\vspace{0mm} \hspace{0mm} {\hbox{\epsfxsize=75mm \epsffile{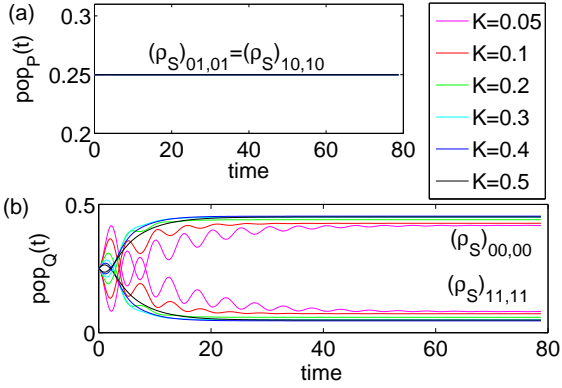}}}
\caption{Population dynamics (a) in the single excitation subspace
$P$ and (b) in the zero and double excitation subspace $Q$. We use
the Bell states (\ref{eq:Bell}) as the initial density matrix with
$a=1/2$. $K$=0.05, 0.1, 0.2, 0.3, 0.4, 0.5, bottom to top; the data
at $K$=0.05 is the most oscillatory. Other parameters are
$\epsilon_{1}=\epsilon_2=0.2$, $\delta=0.1$, $\gamma=0.5$, $T=0.2$
and $J=1$. QUAPI was used with a time step $\delta t=0.25$ and a
memory time $\tau_c=9\delta t$. Convergence was verified by studying
the behavior at different time step $\delta t$ and memory size
$\tau_c$.
 } \label{Fig1}
\end{figure}

\begin{figure}[htbp]
\vspace{0mm} \hspace{0mm} {\hbox{\epsfxsize=80mm \epsffile{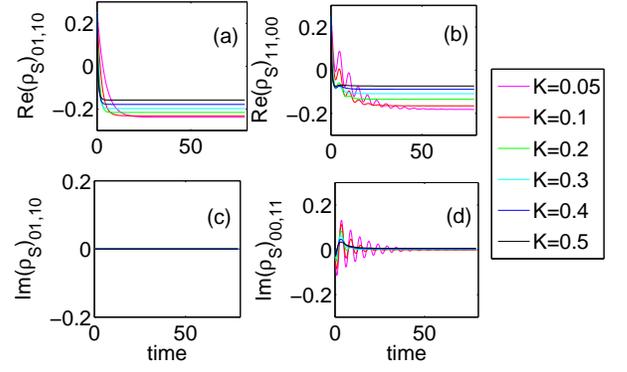}}}
\caption{Real and imaginary parts of the coherences in the single
excitation subspace $P$, (a) and  (c),
and in the zero and double excitation subspace $Q$, (b) and (d).
The different lines were calculated with $K$=0.05, 0.1, 0.2, 0.3, 0.4, 0.5.
Parameters are the same as in Fig. \ref{Fig1}.} \label{Fig2}
\end{figure}

\begin{figure}[htbp]
\vspace{0mm} \hspace{0mm} {\hbox{\epsfxsize=75mm \epsffile{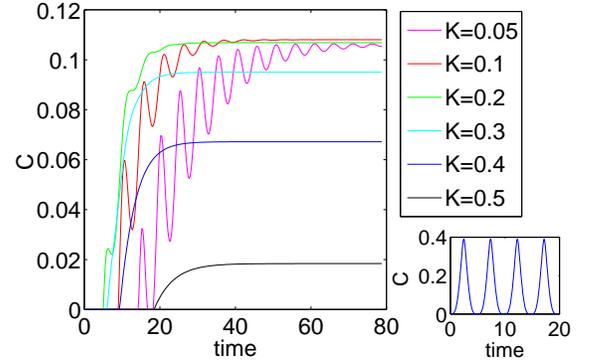}}}
\caption{Concurrence between the two qubits as a function of time,
 manifesting a steady-state
bath-induced entanglement generation. The different lines were
calculated with $K$=0.05, 0.1, 0.2, 0.3, 0.4, 0.5. Parameters are
the same as in Fig. \ref{Fig1}. Right panel: The concurrence
dynamics in the absence of a thermal environment for the same set of
qubits parameters. } \label{Fig3}
\end{figure}

The qubits population behavior in time is displayed in Fig.
\ref{Fig1}, using $a=\frac{1}{2}$.
The qubits have the same energy gap, thus in the $P$ subspace the
two states are degenerate and their population is identical at all
times, independently of $K$. In contrast, in the $Q$ subspace the
energy difference between the states is significant, larger than the
temperature, $T/2\epsilon<1$; the tunneling element is given by
$\gamma J=\frac{1}{2}$, with $\gamma$ as the anisotropy in the
qubit-qubit coupling. In such a situation we expect the steady-state
population of the spin-up state to be significantly smaller than the
ground state population, as indeed we observe in Fig. \ref{Fig1}(b).
An interesting observation is the phenomenon of population inversion
between the zero and the double excitation states before
steady-state sets in. This behavior occurs roughly up to a timescale
that is inversely proportional to the Kondo parameter $K$,
independent of the temperature.
For the same set of parameters Fig. \ref{Fig2} presents the
coherence dynamics in the two subspaces. Generally, coherences are
diminishing with the increase of $K$. Given the population and
coherence dynamics, we display in Fig. \ref{Fig3} the concurrence,
calculated using Eq. (\ref{eq:Conc2}), manifesting a rich dynamics.
The following characteristic's are of particular interest: (i) The
birth-time of the concurrence, (ii) its oscillations, (iii) the
occurrence of sudden death and revival, and (iv) the steady-state
value. We now explain those properties.

{\it Time-zero concurrence.} The particular initial condition used
here, $a=\frac{1}{2}$, results in $C(t=0)=0$. This is because while
we are using maximally entangled states within each subspace as an
initial condition, the entanglement between the two qubits
themselves is zero initially, since all relevant reduced density
matrix elements, necessary for evaluating Eq. (\ref{eq:Conc2}), are
identical \cite{Alber}.

{\it Delayed sudden birth.} When $(\rho_S)_{00,00}\sim
(\rho_{S})_{11,11}$, a situation taking place at, and close to, the
initial time, the concurrence should be zero, given the positivity
condition that limits the value of off-diagonal elements. For small
$K$, the time it takes the system to depart from its initial-equal
population state is prolonged compared to a large-$K$ case, thus,
the concurrence birth-time is delayed with respect to the large $K$
behavior. Interestingly, the delay in the birth time does not extend
linearly with $K$. Rather, the delay is significant for both
$K=0.05$ (weak system-bath coupling) and for $K=0.5$ (intermediate
coupling), while it is shorter for in-between values, $K\sim 0.3$;
The reason is that the delay time is a nontrivial function of both
the time it takes the coherences to establish, and the time it takes
the population to significantly depart from the initial (equally
populated) setup.

{\it Oscillations.} The oscillatory nature of $C$ in time, best
manifested for $K\leq0.1$ reflects the Rabi-type oscillations of the
diagonal elements $(\rho_S)_{00,00}$ and $ (\rho_{S})_{11,11}$. When
these elements are similar in value, the concurrence drops, and even
dies during a certain time interval, depending on the magnitude of
the coherence at that time.

{\it Steady-state value.} If the two qubits are isolated from
thermal effects ($K$=0, right panel of Fig \ref{Fig3}), the
concurrence oscillations reflect the nature of the population and
coherences dynamics, depicting Rabi oscillations. The qubits
behavior under a dissipative thermal bath is notably distinct: Since
both population and coherences approach a constant at long time, the
concurrence reaches a steady-state value as well. It predominantly
reflects the magnitude of the coherence $(\rho_S)_{10,01}$ in the
long time limit since the population weakly depends on $K$ at long
time, see Fig. \ref{Fig1}(b). Interestingly, for the present
$a=\frac{1}{2}$ case the steady-state value of the concurrence is
almost identical at weak system-bath coupling, $K=0.05-0.3$. It
significantly degrades around $K=0.4-0.5$. Beyond that, it is
identically zero.


{\it Sudden death and revival.} Based on Figs.
\ref{Fig1}-\ref{Fig3}, we can draw general conclusions regarding the
process of entanglement sudden death. The effect is directly linked
to the existence of population in the zero and double excitation
states. If the dynamics within the $Q$ space is eliminated all
together ($a=0$), the concurrence is only controlled by the absolute
value of the coherence $|(\rho_S)_{01,10}|$, Eq. (\ref{eq:Conc3}).
This quantity does not manifest an oscillatory behavior: Under the
present initial condition it starts at a large value, touches zero
at a particular time, then grows again to a certain extent.
(Under a different initial condition, e.g.,
$|(\rho_S)_{01,10}(0)|=0$, the entanglement will systematically
grow, up to the steady-state value). In contrast, when the double
excitation state is initially populated, oscillations between the
states in the $Q$ subspace largely occur, if system-bath coupling is
weak. Then, the competition between the two terms in $F_2$, see Eq.
(\ref{eq:Conc1}), can result in a disentanglement over a finite time
interval.

\begin{figure}[htbp]
\vspace{0mm} \hspace{0mm} {\hbox{\epsfxsize=75mm \epsffile{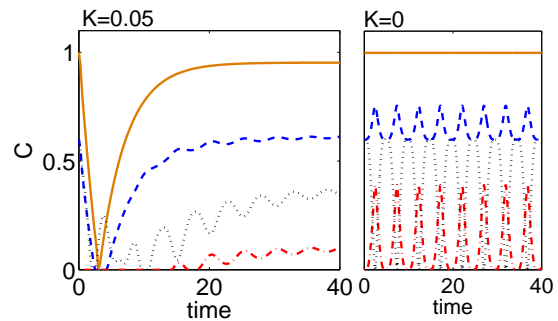}}}
\caption{Concurrence between the two qubits as a function of time,
using Bell states [Eq. (\ref{eq:Bell}] with $a=0$ (full), $a=0.2$
(dashed), $a=0.5$ (dashed-dotted) and $a=0.8$ (dotted). Left Panel:
$K=0.05$. Right Panel: $K=0$. Other parameters are the same as in
Fig. \ref{Fig1}. } \label{Fig4}
\end{figure}

\begin{figure}[htbp]
\vspace{-1mm} \hspace{0mm} {\hbox{\epsfxsize=70mm \epsffile{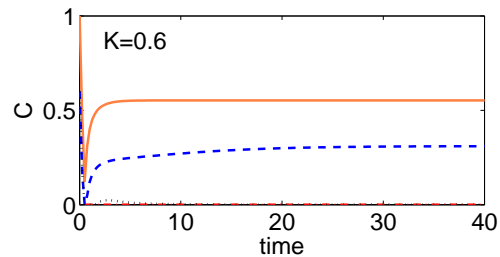}}}
\vspace{-18mm} \caption{Same as Fig. \ref{Fig4} but at strong
system-bath coupling $K=0.6$, $a=0$ (full), $a=0.2$ (dashed),
$a=0.5$ (dashed-dotted) and $a=0.8$ (dotted).
 } \label{Fig4b}
\end{figure}

Fig. \ref{Fig4} displays the concurrence using different initial
conditions, by playing with the parameter $a$. This modifies the
weight of the zero and double excitation states in the dynamics.
When $a=0$ and $K=0$ the entanglement is maximal ($C=1$) at all
times. For finite $K$, keeping $a=0$, it dies at the particular
point at which $|(\rho_S)_{01,10}|=0$. Beyond that, it recovers to a
value close to 1. When we include the $Q$ states, e.g., by taking
$a=0.2$, we observe the effect of entanglement sudden death, over a
certain time interval. The duration of this interval grows when $a$
is further increased up to $a\leq1/2$. Beyond this point the
coherence in the $P$ subspace may dominate over the population in
the $Q$ subspace, resulting in a positive value for $F_1,$
eliminating entanglement sudden death. The behavior at
intermediate-strong system-bath coupling, $K=0.6$, is included in
Fig. \ref{Fig4b}, demonstrating that temporal oscillations are
washed out. The dynamics at even larger $K$ is similar in trends,
with reduced concurrence value.

\begin{figure}[htbp]
\vspace{0mm} \hspace{0mm} {\hbox{\epsfxsize=65mm \epsffile{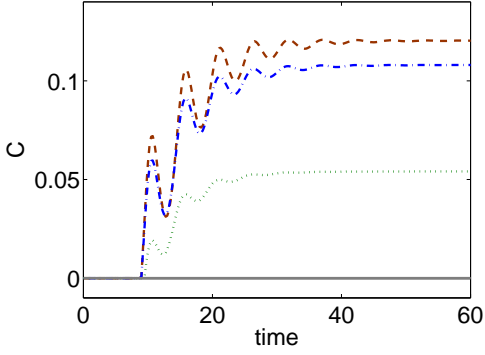}}}
\caption{The role of the bath temperature on the concurrence
evolution. $T=0.1$ (dashed line), $T=0.2$ (dashed-dotted line),
$T=0.4$ (dotted line) and $T=0.6$ (full line). We use Bell states
[Eq. (\ref{eq:Bell})] with  $a=0.5$ and $K=0.1$. Other parameters
are as in Fig. \ref{Fig1}.} \label{Fig5}
\end{figure}

The role of the temperature is displayed in Fig. \ref{Fig5}. At high
temperature the concurrence is zero. At intermediate values,
$T<\epsilon\sim 1$ we find that its sole effect is a shift down of
the qubit entanglement with increasing temperature. All other
features (birth time, oscillation) stay intact. The simulation could
not be performed at temperatures below $T\sim 0.1$ due to
convergence issues in QUAPI.

We can readily study the concurrence under different initial
conditions for the $P$ and $Q$ subspaces, not necessarily in the
form of Bell states, as long as Eq. (\ref{eq:t0}) is obeyed. In
particular, using a diagonal state for the time-zero reduced density
matrix, similar features as those discussed above were obtained.

\section{Universal features at long time}

\begin{figure}[htbp]
\vspace{0mm} \hspace{0mm} {\hbox{\epsfxsize=75mm \epsffile{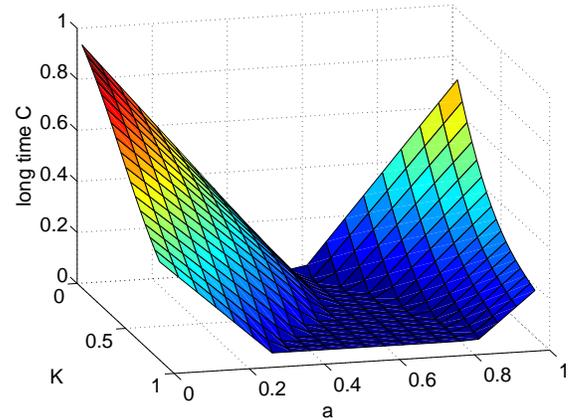}}}
\caption{The long time concurrence representing equilibrium
behavior, for different initial states and system-bath coupling
parameters. $T=0.2$, $J=1$, $\gamma=0.5$, $\delta=0.1$ and
$\epsilon_1=\epsilon_2=0.2$. The long time limit was taken here as
$t=100$. } \label{FigC}
\end{figure}

The long time behavior of the concurrence, representing the
equilibrium limit, is displayed in Fig. \ref{FigC} as a function of
both $K$, the system-bath coupling parameter, and the initial state
preparation ratio $a$, see Eq. (\ref{eq:Bell}). We note that the
concurrence can be significant in both the weak and strong coupling
regimes, as long as the system evolves predominantly in either the
$P$ or $Q$ subspaces. We now show that at weak coupling, $K\ll1$, and at low
temperatures, $T<J\gamma$,
for a broad range of parameters (as we explain below), the following
general result holds
\bea
C_{a<\frac{1}{2}}(t\rightarrow \infty)\sim 1-2a.
\label{eq:Clong0}
\eea
The important implication of this result is that to the lowest order
in $K$ the concurrence deviates from unity due to the occupation of
the zero and doubly excited states in the system. This trend was
observed before, e.g., in Ref. \cite{Zubairy}. However, here, for
the first time, it is justified analytically, based on the
spin-boson model behavior \cite{Weiss}. We derive Eq.
(\ref{eq:Clong0}) by studying the long-time limit of $F_1$, as it
dictates the concurrence when $a\leq \frac{1}{2}$, see Eq.
(\ref{eq:Conc2}). In the biased case, weak coupling theory (beyond
the noninteracting blip approximation) provides  \cite{Weiss}
\bea
\langle Q_z\rangle=(\rho_S)_{00,00}-(\rho_S)_{11,11}\sim a
\frac{\epsilon}{\Delta_b}\tanh\left (\frac{\Delta_b}{T} \right),
\eea
in the thermodynamic limit. Here
 $\Delta_b^2=\epsilon^2+\Delta_{eff}^2$, $\Delta_{eff}$ is a nontrivial
function of $K$, $\omega_c$, and the bare tunneling element in the
$Q$ subspace, $J\gamma$. In the weak coupling limit we can write
$\Delta_{eff}\sim J\gamma$, thus $\Delta_b \sim
\sqrt{\epsilon^2+J^2\gamma^2}$. Manipulating the polarization, we
obtain the relevant term,
\bea
\sqrt{(\rho_S)_{00,00}(\rho_S)_{11,11}}
\sim
\frac{a}{2}\sqrt{ 1-\left( \frac{\epsilon}{\Delta_b} \right)^2
 \tanh^2 \left( \frac{\Delta_b}{T}\right) }.
\label{eq:1}
\eea
The other element in $F_1$ is the coherence in the $P$ subspace.
In the long time limit it satisfies \cite{Weiss}
\bea
|(\rho_S)_{01,10}|\sim
\frac{1-a}{2}\frac{J}{\Omega}\tanh\left(\frac{\Omega}{T}\right),
\label{eq:2}
\eea
where $\Omega^2=J^2 + 2 J^2  K\mu$; the proportionality factor
obeys $\mu= \psi(iJ/\pi T)-\ln(J/T)$ with $\psi$ as the digamma function \cite{Weiss}.
As expected, the equilibrium concurrence depends on the environmental temperature, leading
to entanglement degradation at high $T$, as observed in Fig. \ref{Fig5}.
Considering the low temperature case, $T<J,J\gamma$, we note that
the trigonometric term in both
Eq. (\ref{eq:1}) and Eq. (\ref{eq:2}) is close to unity.
If we further work in the region  $\epsilon <J\gamma$, the square root expression
in Eq. (\ref{eq:1}) gets close to 1.
Under these broad conditions, the concurrence reduces to
\bea
C_{a<\frac{1}{2}}(t\rightarrow \infty)&\sim& (1-a) \frac{1}{\sqrt{1+2\mu K}} - a
\nonumber\\
&\sim&
1-2a-\mu K(1-a).
\label{eq:Clong1}
\eea
One should note that the $K$ dependence is more subtle than the simple linear scale attained here,
since the tunneling element $J$ should be corrected by $K$ in a nontrivial manner \cite{Weiss}.
The simple result (\ref{eq:Clong1}) provides us with some
basic-interesting rules for building a long-time concurrence within the range of parameters
mentioned above:
(i) It decays linearly with the overall population placed
in the $Q$ (zero and double excitation) subspace.
(ii) The reservoir temperature does not significantly affect it.
(iii) It does not depend on the qubit interaction energy.
We note again that these observations are  valid for $a<0.5$,
as long as $T< J,J\gamma$, $\epsilon<J\gamma$ and $K\ll1$.
When $a>0.5$, the concurrence is determined by the competition between $F_1$ and $F_2$, see Eq.
(\ref{eq:Ct}).
The numerics then suggests that $C_{a>\frac{1}{2}}(t\rightarrow \infty)\sim 2a-1$ holds,
 for similar energy parameters.

We conclude this section by emphasizing
the implication of Eq. (\ref{eq:Clong0}):
One could {\it set} the steady-state entanglement in a {\it dissipative} system,
by controlling the initial population in the $P$ and $Q$ subspaces.


\section{Conclusions}

Using exact numerical tools, we simulated the time evolution of two
qubits immersed in thermal environments, considering a class of
initial states for the subsystem. This task was achieved by reducing
the two qubits-bath model into two spin-bath systems, whose
dynamics could be readily followed separately. Using Wootters'
formula for the concurrence \cite{Wooters}, we quantified the degree
of the qubits entanglement in time, exposing rich dynamics,
including oscillations, delayed sudden birth, sudden death, and
revival. Specifically, we showed that the occurrence of entanglement
sudden death can be traced down to the initial population of the
zero and double excitation states. The steady-state behavior was
discussed in the weak coupling limit.

Our results are significant for several reasons. First, we exposed a
general mapping between an interacting two-qubit system embedded in
a bath, and two spin-bath models, allowing us to simulate the
dynamics of the original model using a numerically exact method that
was developed for studying the prominent spin-boson case. Second,
based on our mapping scheme, we calculated the concurrence measure
and demonstrated the essential role of the environment in generating
a stationary entanglement between the (interacting) qubits. By
engineering the environment and its interaction with the system one
could tune the degree of disentanglement in the system \cite{Cirac}.
Earlier studies in this field have mostly treated a simpler version
of our model, ignoring qubit-qubit interaction energy, further
utilizing perturbative treatments. To the best of our knowledge, our
work is the first to calculate the concurrence exactly in a
dissipative and interacting qubit model.

Future studies will focus on the dynamics of non-classical
correlations beyond the entanglement measure, evaluating quantum
discord \cite{Discord1}. This could be done by relying on the
$X$-form  of the reduced density matrix \cite{Alber,Fei}. With this
at hand, we plan to study the dynamics of classical and quantum
correlations in the qubit system, specifically, to investigate
classical and quantum decoherence mechanisms and the possible
transition between them \cite{decoh}.


\begin{acknowledgments}
L.-A. Wu has been supported by the Ikerbasque Foundation Start-up,
the CQIQC grant and the Spanish MEC (Project No.
FIS2009-12773-C02-02) D. Segal acknowledges support from NSERC
discovery grant. The research of C. X. Yu is supported by the Early
Research Award of D. Segal.
\end{acknowledgments}



\end{document}